\documentclass[a4paper]{article}

\usepackage{INTERSPEECH2022}
\usepackage[activate]{microtype}
\usepackage{cite}
\usepackage{color}

\title{Conformer with dual-mode chunked attention for joint online and offline ASR}
\name{Felix Weninger, Marco Gaudesi, Md Akmal Haidar, Nicola Ferri, Jesús Andrés-Ferrer, Puming Zhan}
\address{Nuance Communications, Inc.}
\email{felix.weninger@nuance.com}

\begin{document}

\maketitle
\begin{abstract}
In this paper, we present an in-depth study on online attention mechanisms and distillation techniques for dual-mode (i.e., joint online and offline) ASR using the Conformer Transducer.
In the dual-mode Conformer Transducer model, layers can function in online or offline mode while sharing parameters, and in-place knowledge distillation from offline to online mode is applied in training to improve online accuracy. 
In our study, we first demonstrate accuracy improvements from using chunked attention in the Conformer encoder compared to autoregressive attention with and without lookahead.
Furthermore, we explore the efficient KLD and 1-best KLD losses with different shifts between online and offline outputs in the knowledge distillation.
Finally, we show that a simplified dual-mode Conformer that only has mode-specific self-attention performs equally well as the one also having mode-specific convolutions and normalization. 
Our experiments are based on two very different datasets: the Librispeech task and an internal corpus of medical conversations.
Results show that the proposed dual-mode system using chunked attention yields 5\,\% and 4\,\% relative WER improvement on the Librispeech and medical tasks, compared to the dual-mode system using autoregressive attention with similar average lookahead.
\end{abstract}

\section{Introduction}
End-to-end (E2E) systems have become dominant in automatic speech recognition (ASR) because of their simplicity and better performance. 
In addition to new advancements in model architectures \cite{Han2020-CIC,Gulati2020-CCA,Weninger2021-DEA},
one of the major efforts is to make E2E ASR systems support online streaming applications with strict latency requirements. 
The Recurrent Neural Network Transducer (RNN-T)  has been a favorite E2E model architecture for online streaming because of its time-synchronous processing of input audio and superior performance over the CTC model \cite{Graves2012-STW,Battenberg2017-ENT,Prabhavalkar2017-ACO}.

There have been significant improvements to RNN-T since it was proposed in \cite{Graves2012-STW}, such as replacing the LSTM/BLSTM encoder with Transformer \cite{Yeh2019-TTE,Zhang2020-TTA}, Conformer \cite{Gulati2020-CCA}, and ContextNet \cite{Han2020-CIC}. 
%
A major difference between online streaming and offline batch-mode E2E model is that the former is subject to strict and often application-dependent latency constraints. 
To reduce the deterministic latency incurred during inference, an online ASR system is only allowed to access limited future context.
Since many popular E2E ASR systems are based on bidirectional long-range context modeling (BLSTM, Transformer, etc.) in the encoder, this is the primary reason that online E2E ASR systems generally underperform their offline counterparts. 
The degradation in accuracy is largely determined by the accessed amount of future context. 
There has been extensive research in effectively utilizing future context with limited latency for improving online E2E model performance \cite{Yeh2019-TTE,Zhang2020-TTA,Li2020-OTC,Li2021-ABA,ChunkedMicrosoft,EMFormer}.

From a deployment efficiency point of view, it is beneficial to have a single model able to serve multiple different applications: from offline batch-mode to online streaming under different latency requirements.
Unfortunately, a model trained for the offline use case generally does not perform well in the online use case and vice versa. 
Therefore, there is a direction of research towards making a single model suitable for multiple use cases with different latency requirements \cite{Tripathi2020-TTO,Gao2020-UAU,Audhkhasi2021-MMA,Yu2021-DMA,Kim2021-MMT}.

\textbf{Contributions of our paper:} We extend the dual-mode ASR work in \cite{Yu2021-DMA} in several aspects that were not covered there or in similar works \cite{Audhkhasi2021-MMA,Kim2021-MMT}:
1. Comprehensive evaluation of different online streaming approaches (i.e., autoregressive and chunked attention) based on Conformer Transducer in dual-mode training on two very different data sets. 
2. Evaluation of different distillation approaches for offline-to-online distillation in dual-mode training and the importance of modeling the output shift between offline and online modes. 
3. Propose a dual-mode model trained with shared convolution (i.e., causal convolution) and normalization layers across modes. 


\section{Methodology}

\subsection{Dual-mode Conformer Transducer}

In our paper, we use end-to-end ASR systems based on the Conformer Transducer (Conf-T) architecture, which combines the concept of the recurrent neural network transducer (RNN-T) \cite{Graves2012-STW} with the Conformer encoder \cite{Zhang2020-TTA,Gulati2020-CCA}.
Each Conformer encoder block consists of feedforward, multi-head self-attention (MHSA) \cite{Vaswani2017-AIA}, and convolution layers.
Following the dual-mode approach \cite{Yu2021-DMA}, a single Conformer Transducer model can operate in both online and offline mode.
In online mode, the outputs of convolutions and attention layers are calculated by masking the weights corresponding to future frames. 
Moreover, online and offline mode use different sets of (batch / layer) normalization parameters (running average statistics and scales / offsets). 
For the convolutions, the alternative approach proposed in our paper is to simply use causal (left) padding everywhere.

Offline and online Transducer outputs are calculated as $z_{\text{on}} = M_{\text{on}}^{\theta^\text{on}}(x)$, $z_{\text{off}} = M_{\text{off}}^{\theta^\text{off}}(x) \in [0,1]^{T \times U \times K}$, where $M$ is the model, $\theta^{\text{on}} = [ \theta; \nu^\text{on} ]$, $\theta^{\text{off}} = [ \theta; \nu^\text{off} ]$ are the online and offline model weights, $\nu^\text{on}$, $\nu^\text{off}$ are the corresponding normalization parameters, and $T$, $U$, $K$ denote \# frames, \# tokens, and vocabulary size.
In dual-mode training \cite{Yu2021-DMA}, the offline and online mode are trained jointly while knowledge transfer is done via in-place distillation from the offline to the online mode. 
More precisely, the following loss is minimized:
\begin{equation}
    \mathcal{L} = \alpha \mathcal{L}_\text{trd}(y^*,z_\text{on}) + \beta \mathcal{L}_\text{trd}(y^*,z_\text{off}) + \gamma \mathcal{L}_\text{dist}(z_\text{off},z_\text{on}) ,
    \label{eq:loss}
\end{equation}
where $\mathcal{L}_\text{trd}$ is the transducer loss \cite{Graves2012-STW}, $\mathcal{L}_\text{dist}$ is a distillation loss (cf.\ Section \ref{sec:dist}), 
$y^*$ are the training labels, and $\alpha, \beta, \gamma \geq 0$ are hyperparameters.

\subsection{Chunked Attention}

\label{sec:chunked_attn}

\begin{figure}
    \centering
    \includegraphics[width=\linewidth]{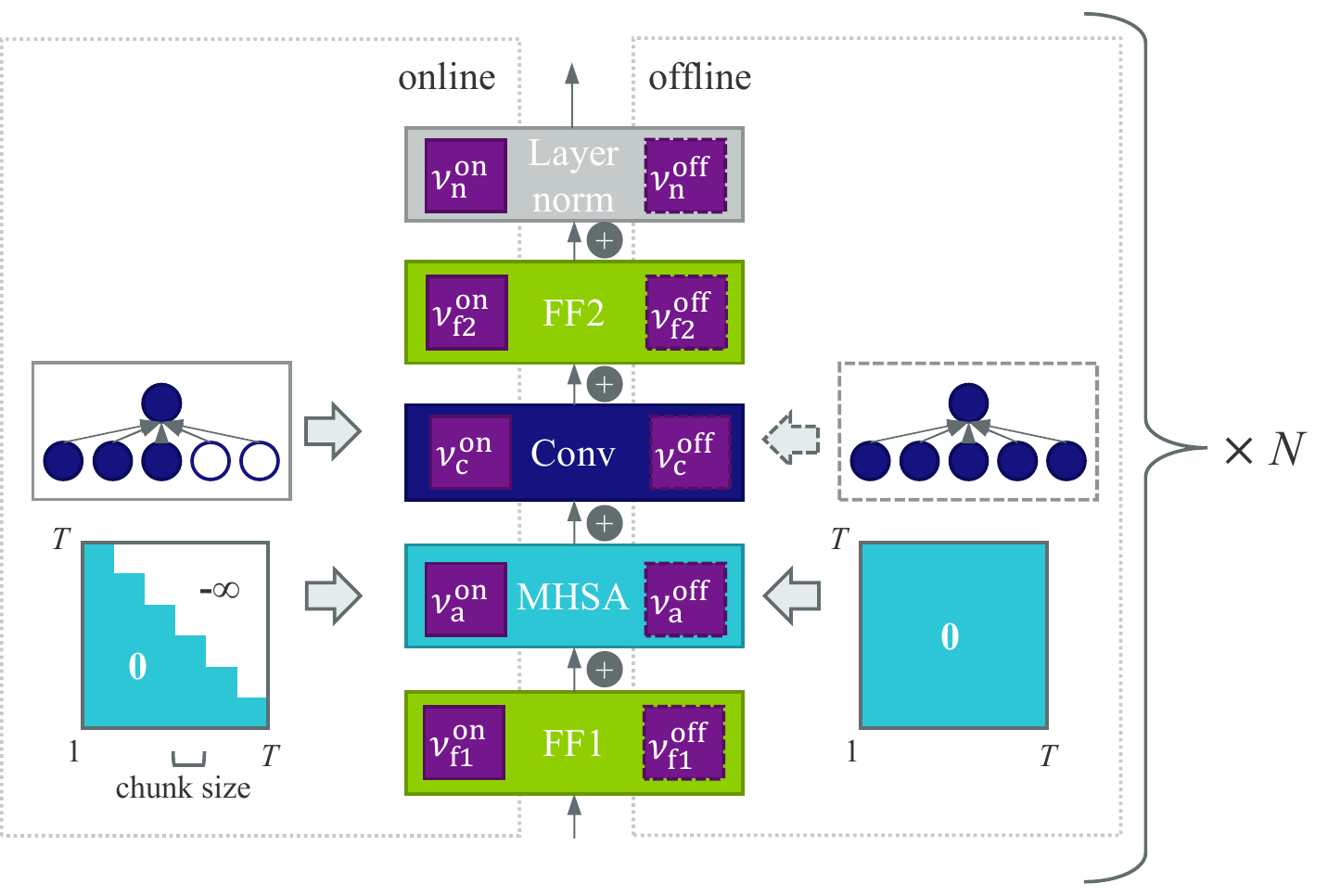}
    \caption{Schema of the dual-mode Conformer encoder with chunked/global attention mask (added to logits), causal/non-causal convolutions and dual-mode normalization with parameters $\nu^{(\cdot)}_{(\cdot)}$. Dashed lines indicate optional components. 
    }
    \label{fig:dual_mode_conformer_chunked_attn}
\end{figure}

To adapt Transformer-like architectures for the streaming use case, the key part to consider is the MHSA block.
The MHSA can be made strictly online by using autoregressive attention \cite{Vaswani2017-AIA}, i.e., every frame in the encoder can only attend to previous frames.
This constraint is efficiently implemented by adding $-\infty$ (or a large negative value) to the attention logits at the `invalid' positions. 
There are several ways to modify this under specified latency requirements, such as truncated lookahead~\cite{Zhang2020-TTA} or contextual lookahead~\cite{EMFormer}. 
While the first approach builds lookahead that accumulates through layers into a larger overall lookahead of the encoder, the latter increases the computational cost at inference by overlapping the input audio chunks. 
In this work, we focus on the chunked attention approach~\cite{ChunkedMicrosoft}, where we divide the input audio into non-overlapping chunks. 
For each encoder input chunk, the MHSA query at each position uses as memory (key and values) all the other positions that belong to the same chunk or previous chunks.
Figure~\ref{fig:dual_mode_conformer_chunked_attn} shows the dual-mode Conformer encoder with chunked attention mask.

\subsection{Distillation}

\label{sec:dist}

In addition to architectural advancements and various approaches for effectively leveraging limited future context, knowledge distillation from offline to online model is another way of improving online model performance \cite{Kurata2020-KDF}. 
We compare the approaches proposed in \cite{Panchapagesan2021-EKD, Yang2021-KDF} in the context of in-place distillation in dual-mode training.

Knowledge distillation by using the Kullback-Leibler divergence (KLD) loss \cite{Yu2013-KDR,Weninger2019-LAS} directly is inefficient for Transducers due to the large size of the output lattice.
The efficient KLD \cite{Panchapagesan2021-EKD} and 1-best distillation \cite{Yang2021-KDF} losses address this issue by restricting the calculation to a reduced output lattice.
Moreover, in order to make these approaches work for offline-to-online distillation, we consider the potential emission delay between online and offline model via a tunable shift parameter $\tau$, similar to \cite{Yu2021-DMA,Yang2021-KDF}.

The efficient KLD loss \cite{Panchapagesan2021-EKD} collapses the probability distribution of the tokens as follows: 
\begin{equation}
    \mathcal{L}_\text{dist}^\text{eff} 
    = \sum_{t,u} \sum_{l \in \{y, \varnothing, r\}} p_\text{off}(l | t,u) \log \frac{p_\text{off}(l | t,u)}{p_\text{on}(l | t-\tau,u)} ,
\end{equation}
where $y$, $\varnothing$ and $r$ denote the correct, the blank, and all other labels, $p_{(\cdot)}$ denotes the probability obtained from the Transducer output $z_{(\cdot)}$, and $t$ and $u$ denote time frame and token indices. 

Conversely, the 1-best distillation loss \cite{Yang2021-KDF} takes into account the full probability distribution, but only along the 1-best path in the teacher lattice.
We extend this approach to in-place distillation by regenerating the 1-best path of the offline model on-the-fly in each training step. 
For consistency, we also use KLD, not cross-entropy as in \cite{Yang2021-KDF}:
\begin{equation}
    \mathcal{L}_\text{dist}^\text{1-best} 
    = \sum_{(t,u) \in \text{1-best}} \sum_{k=1}^K p_\text{off}(k | t, u) \log \frac{p_\text{off}(k | t, u)}{p_\text{on}(k | t-\tau, u)} ,
\end{equation}
where $k$ is the index of a symbol in the vocabulary.

\section{Experiments and Results}

\subsection{Librispeech Data}

\subsubsection{Training recipe}
We first perform a comparative evaluation using the 100 hour training subset of the Librispeech \cite{Panayotov2015-LAA} corpus.
Speed perturbation \cite{Ko2015-AAF} with factors 0.9, 1.0 and 1.1 and SpecAugment \cite{Park2019-SAA} are applied to improve generalization.
The topology of the Conformer Transducer and the training recipe are similar to \cite{Higuchi2021-ACS}. 
The encoder consists of a feature frontend that extracts 80-dimensional log-Mel features, two convolutional layers that perform downsampling on the time axis by a factor of 4, and 18 Conformer blocks with hidden dimension 256 and feed-forward dimension 1024. 
The prediction network has a single LSTM layer with 256 hidden units, and the joint network has 256 units.
The vocabulary contains 30 characters.
Models are trained for 300 epochs.
The training hyperparameters (especially learning rate schedule) were tuned for the offline model using a limited grid search on the clean development set of Librispeech, then applied to all other models (online and dual-mode) without further tuning.
For dual-mode training, the online and offline losses are weighted equally ($\alpha = \beta = 0.5$) and the distillation weight is set to $\gamma=0$ (no distillation) or $\gamma=0.01$.
We measure the word error rate (WER) on the `clean' and `other' test set of Librispeech.
Decoding is done by beam search with beam size 8, without using an external language model.

\subsubsection{Online and offline baselines}

\begin{table}[t]
    \caption{Single-mode baselines on Librispeech 100h (LA: lookahead).}
    \label{tab:results_ls100_single_mode}
    \centering
    \begin{tabular}{l|l|cc}
    \bf Mode & \bf Attention & \multicolumn{2}{c}{\bf Test WER [\%]} \\
     &  & \bf cln & \bf other \\
    \hline
    Online & Autoregressive & 9.6 & 26.9 \\
    Online & Autoreg.\ LA & 8.4 & 24.8 \\
    Online & Chunked & 7.9 & 23.4 \\
    \hline
    Offline (causal conv) & Full context & 6.3 & 18.4 \\
    \quad (non-causal conv) & Full context & 6.3 & 18.3 \\
    Offline \cite{Higuchi2021-ACS} & Full context & 6.8 & 18.9 \\
    \end{tabular}
\end{table}

The results of our single-mode baselines are shown in \tablename~\ref{tab:results_ls100_single_mode}.
Our offline Conformer Transducer system outperforms the reference result obtained by ESPnet \cite{Higuchi2021-ACS}.
%
We also investigated the usage of causal (left padded) 1-D depthwise convolutions in the Conformer blocks in the offline model. 
The WER was similar to the standard non-causal (centered) convolutions. 
Hence, we chose to apply causal convolutions for offline mode as well, thereby simplifying the implementation compared to the original dual-mode Conf-T \cite{Yu2021-DMA}.

For the online systems, we compare autoregressive attention, autoregressive attention with 12 frames ($\approx$ 0.5 seconds) lookahead in the 9th encoder layer\footnote{We did not observe significant performance differences when putting the lookahead in another encoder layer or distributing it across multiple encoder layers.}, and chunked attention (see Section \ref{sec:chunked_attn}) with a chunk size of 25 frames ($\approx$ 1 second).
Using autoregressive attention leads to a drastic WER increase compared to the offline model (52\,\% relative).
However, the relative WER increase is still much smaller than the one reported in \cite{Yu2021-DMA}, suggesting that our online baseline is competitive.
As expected, the lookahead reduces the gap between online and offline WER significantly.
Furthermore, despite having the same average lookahead of about 0.5 seconds, the chunked attention performs better than the autoregressive attention with lookahead (6\,\% WER reduction (WERR)).

\begin{table}[t]
    \caption{Librispeech 100h task: WER obtained by dual-mode systems in online and offline inference with and without efficient KLD distillation (loss weight $\gamma$, shift $\tau$).}
    \label{tab:results_ls100_dual_mode}
    \centering
    \begin{tabular}{l|c|c|cc|cc}
    \bf Online att. & \bf $\gamma$ & $\tau$ & \multicolumn{4}{c}{\bf Test WER [\%]} \\
     & & & \multicolumn{2}{c|}{\bf Online} & \multicolumn{2}{c}{\bf Offline} \\
     &  &  & \bf cln & \bf other & \bf cln & \bf other \\
    \hline
    Autoreg. & 0.0 & -- & 9.0 & 25.2 & 7.2 & 21.8 \\
    Autoreg. & 0.01 & 0 & 9.0 & 25.8 & 7.2 & 21.2 \\
    Autoreg. & 0.01 & -6 & 8.4 & 24.2 & 7.0 & 20.6 \\
    \hline
    Autoreg.\ LA & 0.0 & -- & 7.7 & 23.1 & 6.8 & 19.9 \\
    Autoreg.\ LA & 0.01 & 0 & 7.7 & 22.6 & 7.0 & 19.8 \\
    Autoreg.\ LA & 0.01 & -6 & 7.5 & 22.2 & 6.7 & 19.8 \\
    \hline
    Chunked & 0.0 & -- & 7.4 & 22.0 & 6.4 & 19.2 \\
    Chunked & 0.01 & 0 & 7.7 & 22.4 & 6.4 & 19.2 \\
    Chunked & 0.01 & -6 & \bf 7.1 & 21.5 & \bf 6.1 & 18.9 \\
    \end{tabular}
\end{table}

\subsubsection{Dual-mode systems}

\tablename~\ref{tab:results_ls100_dual_mode} shows the results obtained by dual-mode training for various types of online attention.
Compared to the online baselines in \tablename~\ref{tab:results_ls100_single_mode}, the WER in online mode is improved by dual-mode training in all cases (6\,\%, 8\,\% and 6\,\% WERR on test\_clean for autoregressive, autoregressive with lookahead and chunked attention, respectively).
Furthermore, we observe that there is a consistent gain in online performance from using distillation with shift $\tau=-6$, but no gain from distillation without shift.
Still, the gain from distillation is diminished when lookahead is used, likely because this brings the online performance closer to the offline model and reduces the benefit of knowledge distillation.
The dual-mode system using autoregressive attention in the online mode improves on the WER of the corresponding single-mode online system by 12\,\% relative.
Conversely, the offline performance is degraded by 11\,\% relative.
The trend is similar for the autoregressive attention with lookahead, despite overall better performance.
In contrast, using chunked attention in the online mode avoids the degradation in offline mode, and the corresponding
dual-mode system performs better than the single-mode baselines in both offline and online mode, achieving 10\,\% and 3\,\% relative WERR, respectively.
This is likely because the lookahead for a given frame is not constant in chunked attention,
which makes the online prediction task more similar to the offline one and thus facilitates joint training.

\begin{figure}[t]
    \centering
    \includegraphics[width=\linewidth]{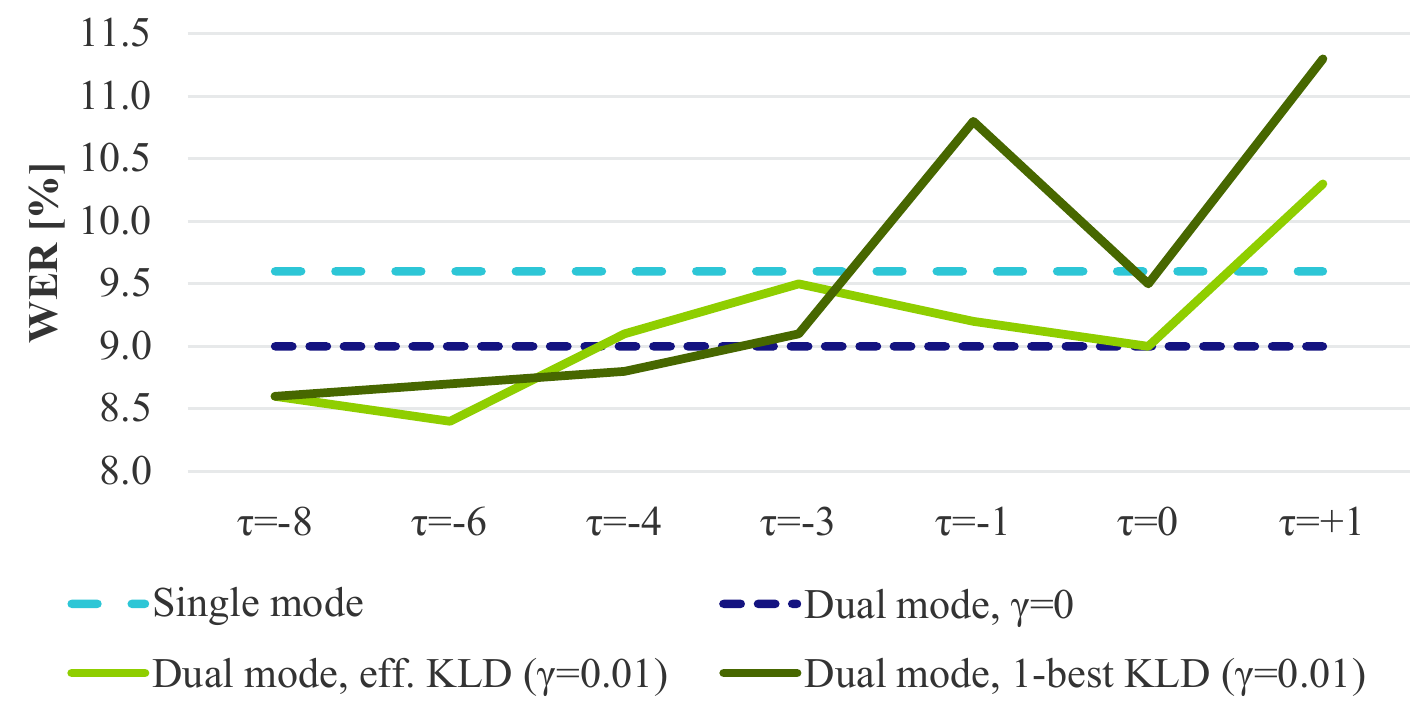}
    \caption{Dual-mode WER on Librispeech 100h task with efficient and 1-best distillation, varying the shift parameter $\tau$.}
    \label{fig:ls100_shift}
\end{figure}

We also investigate the impact of the shift $\tau$ between offline teacher and online student in the in-place distillation with both efficient distillation and 1-best distillation.
As can be seen from \figurename~\ref{fig:ls100_shift}, gains from distillation can be achieved only with fairly large shifts (e.g., $\tau=-6$ corresponds to a $\approx$ 240\,ms emission delay), which is consistent with the findings in \cite{Yang2021-KDF}, while results are unstable for small shifts (in fact, the experiments with $\tau=-2$ diverged).
The best result with efficient distillation is achieved at $\tau=-6$ (8.4\,\% WER), whereas the 1-best distillation performs best with $\tau=-8$.

\subsection{Medical Data}

Additionally, we conduct experiments on an internal data set which consists of conversational speech data in the medical domain (doctor-patient conversations).
The experiments are based on a training set of 1\,k hours manually end-pointed and transcribed speech covering various medical specialties.
We measure WER on a speaker-independent test set consisting of 263\,k words.

The model topology and training recipe are similar to the one used for Librispeech.
In the encoder, we use 16 Conformer blocks with hidden dimension 512 and feed-forward dimension 1024 after the frontend.
The prediction network consists of a single Transformer layer with the same dimensions, and the joint network has 512 units.
The vocabulary contains 2\,k word-pieces.

\begin{table}[t]
    \caption{Single-mode baselines on medical conversation data}
    \label{tab:results_dax_single_mode}
    \centering
    \begin{tabular}{l|l|c}
    \bf Mode & \bf Attention & \bf WER [\%] \\
    \hline
    Online & Autoreg.\ LA &  14.7 \\
    Online & Chunked & 14.4 \\
    \hline
    Offline (causal conv) & Full context & 13.1 \\
    \quad (non-causal conv) & Full context & 13.2 \\
    \end{tabular}    
\end{table}

\tablename~\ref{tab:results_dax_single_mode} shows the single-mode baselines. 
For the online systems, we compare autoregressive attention with lookahead (12 frames) and chunked attention (24 frames).
Unlike on Librispeech, a pure autoregressive model (without any lookahead) did not yield satisfactory performance. 
The chunked attention improves the WER of the online system by 2\,\% relative compared to the autoregressive attention with 12 frames lookahead.
Still, there remains a gap of about 9\,\% relative WER difference between the online and the offline system.

\begin{table}[th]
    \caption{Medical conversation data: WER obtained by dual-mode systems in online and offline inference with and without efficient distillation (loss weight $\gamma$, shift $\tau$).}
    \label{tab:results_dax_dual_mode}
    \centering
    \begin{tabular}{l|c|c|cc}
    \bf Online att. & \bf $\gamma$ & $\tau$ & \multicolumn{2}{c}{\bf WER [\%]} \\
     &  &  & \bf Online & \bf Offline \\
    \hline
    Autoreg.\ LA & 0.0 & -- & 14.2 & 13.3 \\
    Autoreg.\ LA & 0.01 & 0 & 14.2 & 13.3 \\
    Autoreg.\ LA & 0.01 & -6 & 14.1 & 13.2 \\
    \hline
    Chunked & 0.0 & -- & \bf 13.7 & \bf 12.9 \\
    Chunked & 0.01 & 0 & \bf 13.7 & \bf 12.9 \\
    Chunked & 0.01 & -6 & 13.8 & 13.0 \\
    \end{tabular}
\end{table}

\tablename~\ref{tab:results_dax_dual_mode} shows the results obtained by dual-mode training.
We use the efficient KLD loss in case of $\gamma > 0$.
As in the Librispeech scenario, using chunked attention in the online mode helps improving both online and offline performance. The dual-mode system with chunked attention obtains 3.7\,\% / 3.1\,\% relative WERR compared to the one using autoregressive attention with lookahead, and 5.4\,\% / 1.6\,\% with respect to the corresponding single-mode online / offline system.
However, unlike on Librispeech, we do not observe any gain from distillation, even with $\tau=-6$.
One possible reason is that the performance difference  between the online and offline models on the medical data set is small (about 9\,\% relative, see \tablename~\ref{tab:results_ls100_single_mode}) compared to that on the Librispeech data set (23\,\% relative, see \tablename~\ref{tab:results_ls100_single_mode}).

\subsection{Emission Timing}

\begin{figure}[t]
    \centering
    \includegraphics[width=.9\linewidth]{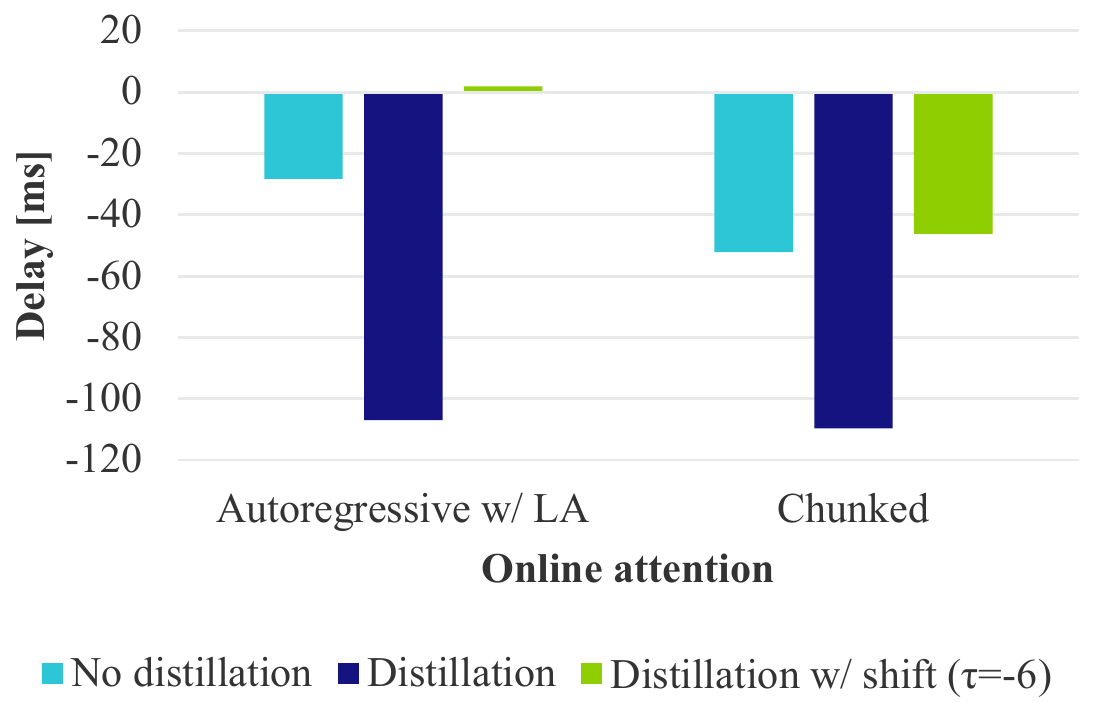}
    \caption{Emission delay of dual-mode systems vs.\ single-mode online reference measured on medical data (lower means earlier).}
    \label{fig:dax_latency}
\end{figure}

\figurename~\ref{fig:dax_latency} shows the time delay of transcriptions produced by our dual-mode models, with respect to a single-mode reference.
We compute this delay as the average time difference of all the matching couples of correct words. For each word we consider the time of the emitting frame of its last word-piece in the RNN-T output alignment. If the encoder is chunked, times are rounded up to the end of their corresponding encoder chunks.
The absolute emission delay is similar between autoregressive attention with lookahead and chunked attention models.

As can be seen in \figurename~\ref{fig:dax_latency}, dual-mode models typically emit faster than the single-mode ones.
When the dual-mode model is trained without distillation, we observe a slightly lower delay in the chunked configuration compared with the autoregressive one, while a significant improvement in terms of emission delay ($\approx$ 110\,ms) is evident in both cases when the distillation is enabled.
However, the latency gain from distillation vanishes when the teacher targets are shifted with a negative $\tau$ value, because this configuration encourages later emission.

\subsection{Effect of Dual-mode Normalization and Joint Training}

In \tablename~\ref{tab:results_ls100h_dax_dual_single_norm}, we assess the importance of the dual-mode normalization layers in the Conformer blocks proposed by \cite{Yu2021-DMA} vs.\ simply sharing the normalization layers between online and offline mode.
On Librispeech (using chunked attention and efficient distillation with $\tau=-6$), the online performance is very similar between dual and single-mode normalization, while there is a small degradation in offline WER.
The medical conversation task (using chunked attention but no distillation) shows a similar picture.
Since we use causal convolutions for both online and offline mode as in the previous experiments, using a single set of normalization layers means that convolutional and feedforward components are identical to the single-mode Conformer, and the MHSA layers vary only the attention mask.
Thus, single-mode normalization considerably simplifies the implementation while yielding similar performance.

\begin{table}[t]
    \caption{WER obtained by dual-mode systems in online and offline inference, using dual normalization layers (one for online and one for offline) or single normalization layers (shared between online and offline mode).}
    \label{tab:results_ls100h_dax_dual_single_norm}
    \centering
    \begin{tabular}{l|c|c}
    \bf Norm.\ layers & \multicolumn{2}{c}{\bf WER [\%]} \\
    & \bf Online & \bf Offline \\
    \hline
    \multicolumn{3}{c}{\em Librispeech 100h (clean / other)} \\
    \hline
    dual & 7.1 / 21.5 & 6.1 / 18.9 \\
    single & 7.1 / 21.3 & 6.3 / 19.0 \\
    \hline
    \multicolumn{3}{c}{\em Medical conversation task} \\
    \hline
    dual & 13.7 & 12.9  \\
    single & 13.7 & 13.0 \\
    \end{tabular}
\end{table}

Motivated by these results, we also investigated a further simplification of the dual-mode training where the attention mask for all MHSA layers is randomly chosen as the global (offline) or chunked (online) one for each line in the current mini-batch, instead of training both modes on the entire batch (joint training, cf.\ Eq.\ \eqref{eq:loss}).
This is similar in spirit to the sampling techniques in \cite{Yu2021-DMA,Audhkhasi2021-MMA,Kim2021-MMT}.
The advantage is that only one model (online or offline) is computed for each utterance, thus saving approximately 50\,\% of computation and memory requirement.
We found such dual-mode training to yield a single model for both online and offline mode that performed similar to the dedicated single-mode models (14.2\,\% / 13.3\,\% WER on the medical task). 
However, unlike joint training, it did not result in a sizable WER gain compared to the single-mode baselines.

\section{Conclusions}

In this paper, we presented an in-depth study on the performance of dual-mode training for online Conformer Transducer architectures. 
We could obtain significant WER improvements in online mode on both the Librispeech and a medical conversational speech task, even without in-place distillation, and match the performance of dedicated offline models.
Best results in online mode were obtained using chunked attention.
Our results also shed light on the importance of modeling emission delay when doing offline-to-online knowledge distillation: we found that distillation without shift is helpful for reducing latency, while distillation with shift can reduce the WER at the expense of emission delay.
The latter could potentially be mitigated by techniques such as FastEmit \cite{Yu2021-FLS}.
In general, the gain from distillation depends on the online configuration (especially the lookahead) and the data set.
Furthermore, we explored several modifications to the original training approach, and found a simplified version, where only the attention mask is exchanged between online and offline modes, 
to perform equally well as the original proposal \cite{Yu2021-DMA}.
In future work, we will apply our findings to multi-mode ASR \cite{Kim2021-MMT} for improving robustness of the online model in multiple latency requirements. 


\bibliographystyle{IEEEtran}

\bibliography{refs}

\end{document}